\documentclass[12pt,reprint,aps,groupedaddress,showpacs,nofootinbib,prd,twocolumn]{revtex4-2}

\usepackage{amsmath,amssymb,graphicx,amsfonts}
\usepackage{mathrsfs}
\usepackage{comment}
\usepackage{xcolor}
\usepackage[shortlabels]{enumitem}
\usepackage{hyperref}
\usepackage{url}

\begin{document}
\title{Dynamical behavior of the $C$-metric: charged scalar fields, quasinormal modes and superradiance}

\author{Kyriakos Destounis$^1$, Giacomo Mascher$^{1,2}$, and Kostas D. Kokkotas$^1$} 
\affiliation{$^1$Theoretical Astrophysics, IAAT, University of T{\"u}bingen, 72076 T{\"u}bingen, Germany}
\affiliation{$^2$Physics Department, University of Trento, Via Sommarive 14, 38123 Trento, Italy}

\begin{abstract}
The $C$-metric is a boost-symmetric spacetime solution to the vacuum Einstein field equations which describes black holes that are uniformly accelerated under the tension of a cosmic string. Only recently the thermodynamics of accelerating black holes and their modal stability against neutral scalar perturbations were concisely established. The generalization of accelerating black holes to incorporate an electric charge, namely the charged $C$-metric, possess three distinct families of quasinormal mode frequencies; the complex photon surface quasinormal modes associated with unstable null particles at the equatorial plane of the photon surface, the purely imaginary acceleration modes whose existence solely depends on the acceleration of spacetime and the purely imaginary near-extremal modes which dominate the dynamics of the ringdown at late times when the event and Cauchy horizon approach each other. We extend the quasinormal mode analysis to charged scalar fluctuations and find that the photon surface modes are continuously deformed with respect to their neutral counterpart as the scalar charge is increased. We further find that the acceleration and near-extremal families acquire an oscillation frequency when the scalar charge is introduced. Finally, we study the superradiant amplification of charged scalar monochromatic waves impinging charged accelerating black holes. We find that even though the frequency range of superradiant amplification is lessened due to the acceleration, the amplification factors are considerably elevated with respect to those transpiring in Reissner-Nordstr\"om black holes and are maximized when the scalar charge is significantly large.
\end{abstract}
	
\maketitle

\section{Introduction}

Black holes (BHs) are unequivocally amongst the most fascinating astrophysical objects of our Cosmos. They arise as exact solutions of the Einstein field equations and serve as perfect testbeds for strong field gravity \cite{Barack:2018yly}. Our current understanding of General Relativity (GR) and the constraints of its modifications stems from BH phenomenology through numerical and observational investigations \cite{Abbott_2017}. In the majority of such scenarios, BHs are considered as isolated objects in vacuum which are barely influenced by their dynamic environment \cite{Barausse:2014tra,Cardoso:2021wlq}, besides accretion. Currently, the Kerr-Newman family provides a four-dimensional prototypical solution of GR, parameterized by its mass, angular momentum and charge, which describes an isolated, topologically spherical compact object.

Yet, there are further esoteric solutions to GR, that describe a special class of boost-symmetric BH geometries which can potentially represent moving compact objects \cite{Cardoso:2019dte}. Among these solutions, the $C$-metric \cite{Weyl} is a special case of Petrov type D, hence it can admit charge \cite{Kinnersley_1970} and rotation \cite{Plebanski:1976gy}. These spacetimes are algebraically general, radiative \cite{Bicak_1968,Farhoosh,Bicak:1989knv,Bicak_1999,Pravda:2000vh} and possess a plausible Newtonian limit \cite{Bicak:2009xt}. Their causal structure can be interpreted as a pair causally disconnected BHs which uniformly accelerate away from each other in opposite directions. The perpetual acceleration is provided by a conical deficit along one polar axis which represents a finite-width cosmic string \cite{Gregory:1995hd} that pulls the event horizon. Cosmic strings do not produce long range local curvature, but generate an overall global conical deficit in the spacetime which provides the driving force of an accelerating BH. Thus, one can interpret the $C$-metric as a BH that has been accelerated under its interaction with a local cosmological medium, such as a cosmic string in tension\footnote{The conical deficit can be replaced by the introduction of a more physical magnetic field, leading to the Ernst spacetime \cite{Ernst_1976,Ernst_1978}. This solution does not require the existence of cosmic strings though has significantly altered non-asymptotically flat structure.}. 

A cosmic string \cite{Kibble:1976sj,Vilenkin:1984ib,Witten:1984eb,Hindmarsh:1994re,Copeland:2009ga} is a one-dimensional topological defect with a stress-energy tensor and a string tension proportional to its mass per unit length. Cosmic strings may have formed during a symmetry-breaking phase transition in the early Universe \cite{Kibble:1980mv,Vachaspati:1984dz}, right after cosmological inflation, and are a generic prediction of quantum field theories \cite{Everett:1981nj} and string theory models \cite{Damour:1996pv,Kachru:2003sx,Copeland:2011dx}. Cosmic string evolution \cite{Albrecht:1984xv,Bennett:1989ak,Allen:1990tv}, their electromagnetic \cite{Vilenkin:1986zz} and gravitational radiation \cite{Caldwell:1991jj,Allen:1991bk}, as well as their impact on the cosmic microwave background radiation \cite{Kaiser:1984iv,Bouchet:1988,Coulson:1993ig,Jeong:2004ut,Battye:2010PhRvD..81l3512B,Landriau:2011,Hergt:2016xup}, pulsar timing \cite{Bouchet:1989ck,Jenet:2006sv}, gravitational lensing \cite{Gott:1985ApJ...288..422G,Mack:2007ae,Thomas:2009bm}, galaxy redshift surveys \cite{Brandenberger:2010hn}, galactic formation \cite{Silk:1984xk,Turok:1985tt,Brandenberger:1987er} and the GW stochastic background \cite{Damour:2000wa,Damour:2004kw,Siemens:2006vk,Siemens:2006yp,Abbott:2009,Olmez:2010bi,Binetruy:2010cc,LIGOScientific:2013tfe} have placed tight constraints on their parameters and various theories that lead to their production. Even though there is currently no observational evidence for the existence of cosmic strings, null results do not necessarily disprove them. 

Beyond classical GR, the $C$-metric has been utilized to analyze the splitting of cosmic strings \cite{Eardley:1995au}, BH pair creation \cite{Hawking:1995zn,Hawking-Horowitz-Ross,Hawking:1997ia} and in the construction of five-dimensional black rings \cite{Emparan:2001wn}. Nevertheless, accelerating BHs are not extensively used since their thermodynamics was only recently established \cite{Appels_2016,Anabalon:2018qfv,Anabalon:2018ydc,Gregory_2019}.

To place the $C$-metric in the arsenal of BH practitioners, their stability against perturbations is a prerequisite in order for further investigations to take place. The separability of test field master equations \cite{Kofron_2015,Bini:2014kga} and the quasinormal modes (QNMs) of the charged $C$-metric \cite{Destounis:2020pjk} designate that such BHs are modally stable against neutral scalar fields, since their QNM frequencies, which characterize the BH's ringdown signal, decay monotonously in time. Furthermore, the Cauchy horizon of near-extremally charged accelerating BHs has been shown to be sufficiently stable \cite{Destounis:2020yav}, thus violating the strong cosmic censorship conjecture \cite{Cardoso:2017soq}, while the shadow of accelerating rotating BHs has been found to be significantly distorted with respect to that of a Kerr BH \cite{Zhang:2020xub}.

In this article, we analyze the stability of charged accelerating BHs against electrically charged scalar fluctuations, thus extending current results which regard neutral scalar fields \cite{Destounis:2020pjk}. We find that the spacetime remains modally stable by calculating the respective charged scalar QNMs and discuss their behavior in the complex plane when the charge coupling between the scalar and BH charge is introduced. Moreover, due to the similarities of charged accelerating and Reissner-Nordstr\"om-de Sitter (RNdS) BHs \cite{Destounis:2020pjk,Destounis:2020yav} as well as charged BHs surrounded by anisotropic fluids \cite{Cuadros-Melgar:2021sjy}, we initiate the study of charged-scalar-wave scattering of such compact objects and investigate the occurrence of superradiance \cite{Zeldovich,Penrose:1971uk,Bekenstein:1973mi,Brito:2015oca}; a process through which a wave scattered off a BH is amplified under the expense of the BH's energy. We identify the appropriate frequency range for which superradiant amplification of charged scalar monochromatic waves occurs and find that the introduction of acceleration further augments superradiance with respect to the amplification factors found for charged scalar waves impinging Reissner-Nordstr\"om (RN) BHs.

In what follows, we employ the geometrized unit system such that the gravitational constant and speed of light equal to unity, that is $G=c=1$.

\section{The charged $C$-metric}

The charged $C$-metric is a solution to the Einstein-Maxwell field equations and generalizes the RN geometry to include an additional parameter, besides the mass $M$ and electric charge $Q$; the BH's acceleration parameter $\alpha$. The line element can be written in spherical-type coordinates (although the spacetime is not spherically-symmetric) so that it covers one of the charged accelerating BHs, as \cite{Griffiths:2006tk,griffiths_podolsky_2009}
\begin{align}
	\label{Cmetric}
	\nonumber
	ds^2=\Omega^{-2}\left(-f(r)dt^2+f(r)^{-1}dr^2+P(\theta)^{-1} r^2 d\theta^2\right.\\
	\left.
	+P(\theta)r^2 \sin^2\theta d\varphi^2\right),
\end{align}
where
\begin{align}\label{conf}
	\Omega(r,\theta)&=\left(1-\alpha r \cos\theta\right),\\
	f(r)&=\left(1-\frac{2M}{r}+\frac{Q^2}{r^2}\right)(1-\alpha^2 r^2),\\
	P(\theta)&=1-2\alpha M \cos\theta+\alpha^2 Q^2 \cos^2\theta.
\end{align}
The electromagnetic potential associated with the charged BH source is given by $A_\mu=(-Q/r,0,0,0)$. The metric \eqref{Cmetric} asymptotes to the RN solution as $\alpha\rightarrow 0$ and to the $C$-metric when $Q\rightarrow 0$. There is a curvature singularity at $r=0$, while the roots of $f(r)$ determine the existence of three null hypersurfaces (see Fig. 1 in \cite{Destounis:2020pjk}) at 
\begin{eqnarray}
	&&r=r_\alpha:=\alpha^{-1},\\
	&&r=r_{\pm}:=M\pm\sqrt{M^2-Q^2}, 
\end{eqnarray}
namely the acceleration horizon $r=r_\alpha$, event horizon $r=r_+$ and Cauchy horizon $r=r_-$ radius, which satisfy the inequality $r_-\leq r_+\leq r_\alpha$. Our analysis will take place at the static region $r_+<r<r_\alpha$, where $f(r)$ is positive and the line element has fixed signature, implying that $P(\theta)>0$ for all $\theta\in\left[0,\pi\right]$. We particularly avoid the region $r\geq r_\alpha$ since the uniformly accelerating reference frame of the BH asymptotically approaches the speed of light at $r=r_\alpha$, thus events beyond the acceleration horizon are causally disconnected to the static region. Since $r_+\leq r_\alpha$ needs to hold, we require $\alpha\leq 1/r_+$, where at the equality the BH becomes extremal and the event and acceleration horizons coincide. Additionally, when $M=Q$, the event and Cauchy horizons meet and the BH is again extremal.

Conical singularities generally occur on the south $\theta=0$ and north pole $\theta=\pi$, due to the fact that the ratio of the circumference over the radius of the object is not exactly $2\pi$ there, designating the existence of excess or deficit angles. By specifying the range of $\varphi$ accordingly, the deficit or excess angle of one of these topological singularities can be removed. By assuming $\varphi\in  [0,2\pi C )$, where $C=1/P(\pi)$ a constant, we can remove the excess angle at $\theta=\pi$ and interpret the metric as a charged BH that is being accelerated along the pole axis $\theta=0$ by a cosmic string. Equivalently, we can remove the deficit angle at $\theta=0$ and obtain an excess angle at $\theta=\pi$ by setting $C=1/P(0)$. In this case, the charged BH is accelerated along the pole axis $\theta=\pi$ by a cosmic strut (see \cite{Griffiths:2006tk,Destounis:2020pjk} for a more detailed discussion).

\section{Decoupled master equations for charged scalar fields}

The charged $C$-metric, described by Eq. \eqref{Cmetric}, can be conformally rescaled by choosing $\Omega$ as in Eq. \eqref{conf} so that
\begin{align}
	\label{comp_Cmetric}
	\nonumber
	d\tilde{s}^2=\Omega^2 ds^2=-f(r)dt^2+f(r)^{-1}dr^2+P(\theta)^{-1} r^2 d\theta^2\\
	+P(\theta)r^2 \sin^2\theta d\varphi^2.
\end{align}
Since the $C$-metric is four-dimensional, the conformal wave equation (see Eq. \eqref{conformal_wave_equation}) is written as 
\begin{equation}\label{conformal_wave_equation_4d}
	g^{\mu\nu}D_\mu D_\nu\phi-\frac{1}{6}R\phi=0,
\end{equation}
where $g^{\mu\nu}$ and $R$ are the metric tensor and the Ricci scalar of Eq. \eqref{Cmetric}, while $D_\mu=\nabla_\mu-iqA_\mu$ and $\phi$ is the scalar field with charge $q$. The $C$-metric \eqref{Cmetric} has been shown to be asymptotically flat \cite{Hawking:1997ia,ashtekar1981}, thus it follows that $R=0$. This means that by solving \eqref{conformal_wave_equation_4d}, we effectively solve the standard Klein-Gordon equation for charged scalar fields 
\begin{equation}\label{wave_eq}
	g^{\mu\nu}D_\mu D_\nu\phi=0.
\end{equation}
Although Eq. \eqref{conformal_wave_equation_4d} is not separable when using the metric \eqref{Cmetric}, it can be shown that by using the conformally-invariant equation 
\begin{equation}\label{conformal_tilde_wave_equation_4d}
	\tilde{g}^{\mu\nu}\tilde{D}_\mu \tilde{D}_\nu\tilde{\phi}-\frac{1}{6}\tilde{R}\tilde{\phi}=0,
\end{equation}
where $\tilde{\phi}=\Omega^{-1}\phi$, $\tilde{g}_{\mu\nu}=\Omega^2 g_{\mu\nu}$ and $\tilde{D}_\mu,\,\tilde{A}_\mu,\, \tilde{R}$ are given in Appendix \ref{appA}, the radial and polar counterparts of the scalar field are separable through the ansatz
\begin{equation}\label{decomposition}
	\tilde{\phi}=e^{-i\omega t}e^{i m\varphi}\frac{\psi(r)}{r}\chi(\theta),
\end{equation}
where $\omega$ is the frequency and $m$ the azimuthal number of the scalar field. Hence, by solving Eq. \eqref{conformal_tilde_wave_equation_4d} we effectively solve Eq. \eqref{conformal_wave_equation_4d} and consequently the wave equation \eqref{wave_eq} which describes the propagation of charged scalar fields (see Appendix \ref{appB} for further details).

By expanding \eqref{conformal_tilde_wave_equation_4d} with the conformally-rescaled spacetime metric \eqref{comp_Cmetric} we obtain
\begin{align}
	\nonumber
	&-\frac{r^2\partial^2_t\tilde{\phi}}{f(r)}+\partial_r(r^2f(r)\partial_r\tilde{\phi})
	-\frac{2iqQ}{f(r)r}\partial_t\tilde{\phi}+\frac{q^2Q^2}{f(r)r^2}\tilde{\phi}\nonumber\\
	&+\frac{1}{\sin\theta}\partial_\theta(P(\theta)\sin\theta\partial_\theta\tilde{\phi})+\frac{\partial^2_\varphi\tilde{\phi}}{\sin^2\theta P(\theta)}\nonumber\\\nonumber
	&+\frac{1}{6}\left(r^2 f^{\prime\prime}(r)+4rf^{\prime}(r)+2f(r)+P^{\prime\prime}(\theta)\right.\\
	&\left.+3\cot\theta P^{\prime}(\theta)-2P(\theta)\right)\tilde{\phi}=0,
	\label{feq1}
\end{align}
where the primes denote differentiation with respect to the function's variable. Since the azimuthal coordinate $\varphi$ is periodic and the conical deficit along $\theta=\pi$ is removed by requiring $C=1/P(\pi)$, then $m$ must have the form $m=m_0 P(\pi)$, with $m_0\geq0$ the principal azimuthal number associated with the angular number $\ell$.

Consequently, by using \eqref{decomposition}, Eq. \eqref{feq1} reduces to two decoupled ordinary differential equations for the radial and polar sectors
\begin{eqnarray}
	\label{final_radial}
	\frac{d^2\psi(r)}{dr^2_*}+\left[\omega^2-2\omega\Phi(r)-V_r\right]\psi(r)&=0,\\
	\label{final_polar}
	\frac{d^2\chi(\theta)}{dz^2}-[m^2-V_\theta]\chi(\theta)&=0,
\end{eqnarray}
where
\begin{equation}
\Phi(r)=\frac{qQ}{r},\,\,\,\,\,\,\,\,\,	dr_*=\frac{dr}{f(r)},\,\,\,\,\,\,\,\,\,dz=\frac{d\theta}{P(\theta)\sin\theta},
\end{equation}
and
\begin{align}
	\label{pot_r}
	V_r&=f(r)\left(\frac{\lambda}{r^2}-\frac{f(r)}{3r^2}+\frac{f^\prime(r)}{3r}-\frac{f^{\prime\prime}(r)}{6}\right)-\Phi(r)^2,\\\nonumber
	V_\theta&=P(\theta)\left(\lambda \sin^2\theta-\frac{P(\theta)\sin^2\theta}{3}+\frac{\sin\theta\cos\theta P^\prime(\theta)}{2}\right.\\
	\label{pot_theta}&\,\,\,\,\,\,\,\,\,\,\,\,\,\,\,\,\,\,\,\,\,\,+\left.\frac{\sin^2\theta P^{\prime\prime}(\theta)}{6}\right),
\end{align}
with $\lambda$ a separation constant.

\section{Quasinormal modes}

In order to solve Eqs. \eqref{final_radial} and \eqref{final_polar}, we need to impose appropriate boundary conditions. For QNMs, we use the following physically-motivated boundary conditions \cite{Hawking:1997ia,Destounis:2020pjk,Destounis:2020yav}
\begin{align}
	\label{bcs_radial}
	\psi(r) &\sim
	\left\{
	\begin{array}{lcl}
		e^{-i\left(\omega-\Phi(r_+)\right) r_* },\, r_*\rightarrow-\infty \,\,(r \rightarrow r_+),\\
		&
		&
		\\
		e^{+i\left(\omega-\Phi(r_\alpha)\right) r_*},\, r_*\rightarrow+\infty\,\,(r \rightarrow r_\alpha),
	\end{array}
	\right.\\
	\label{bcs_polar}
	\chi(\theta) &\sim
	\left\{
	\begin{array}{lcl}
		e^{+mz},\, z\rightarrow-\infty \,\,(\theta \rightarrow 0), \\
		&
		&
		\\
		e^{-mz},\, z\rightarrow+\infty\,\,(\theta \rightarrow \pi).
	\end{array}
	\right.
\end{align}\\
Although the purely ingoing (outgoing) boundary conditions \eqref{bcs_radial} at the event (acceleration) horizon are obvious for QNMs, conditions \eqref{bcs_polar} are taken so that the scalar field solutions do not blow up at the interval boundaries of $\theta$ \cite{Hawking:1997ia,Destounis:2020pjk,Destounis:2020yav}. 

When $\alpha\rightarrow 0$, the separation constant takes the exact form $\lambda=\ell(\ell+1)+1/3$ \cite{Kofron_2015}. Hence, we can map each $\lambda$ obtained by solving Eq. \eqref{final_polar} with boundary conditions \eqref{bcs_polar} to a certain angular number $\ell$, e.g., in the limit $\alpha\rightarrow 0$, the value $\lambda=1/3$ corresponds to $\ell=0$, $\lambda=7/3$ corresponds to $\ell=1$ and so on.

The solutions to the radial equation \eqref{final_radial} with boundary conditions \eqref{bcs_radial}, together with the separation constant calculated previously, leads to a discrete set of QNMs $\omega_n$ depending on the choice of the BH parameters, magnetic quantum number $m$, separation constant $\lambda$ and overtone number $n$, where the $n=0$ mode is the fundamental QNM that dominates the late-time behavior of the ringdown signal \cite{Destounis:2020pjk,Destounis:2020yav}, while $n>0$ modes correspond to overtones.

To calculate QNMs, we utilize the \emph{Mathematica} package \emph{QNMSpectral} developed in \cite{Jansen:2017oag}, which is based on the discretization of differential equations using pseudospectral collocation methods \cite{Dias:2010eu} and directly solves the resulting generalized eigenvalue problem.

\section{Superradiant amplification of charged scalar waves}

\begin{figure*}[]
	\includegraphics[scale=0.4]{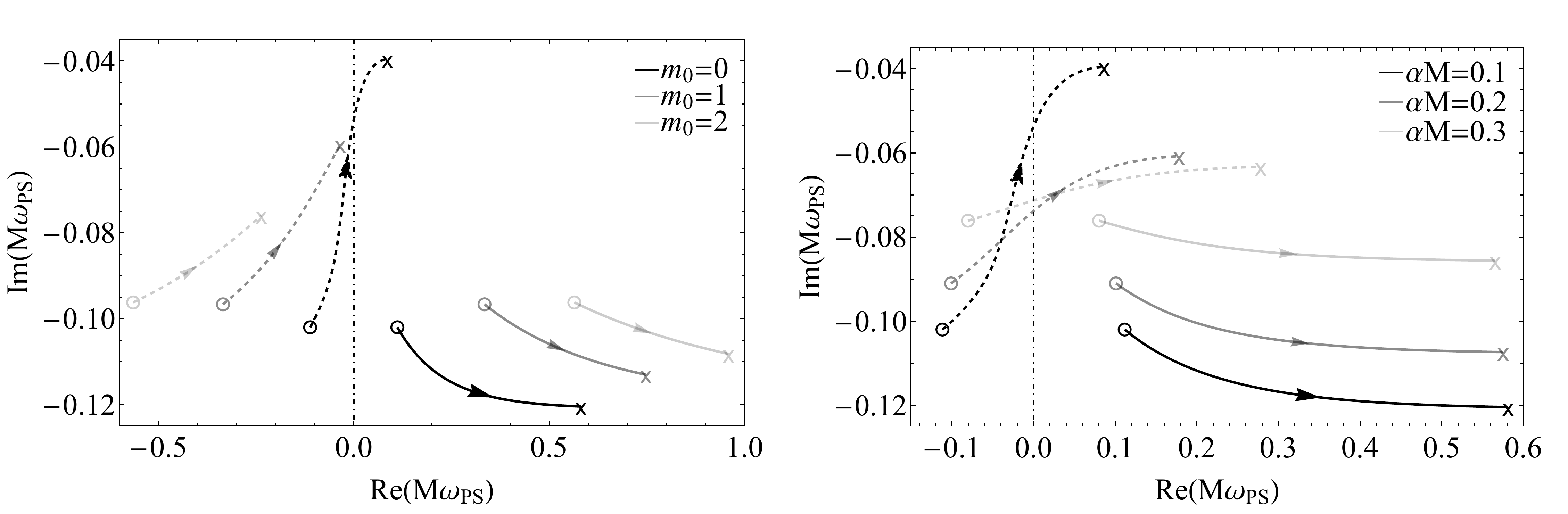}
	\caption{Left: Fundamental ($n=0$) charged scalar PS QNMs for an accelerating RN BH with $Q=0.5M$, $\alpha M=0.1$ and varying principal azimuthal number $m_0$. The charge coupling $qQ$ ranges from $0$ (circles) to $1$ (cross marks). Right: Fundamental ($n=0$) charged scalar PS QNMs with $m_0=0$ for an accelerating RN BH with $Q=0.5M$ and varying acceleration $\alpha M$. The charge coupling $qQ$ ranges from $0$ (circles) to $1$ (cross marks) following the direction of the arrows.}
	\label{PSQNMs}
\end{figure*}

To explore the superradiant amplification of charged monochromatic incident waves scattering off the charged $C$-metric's photon surface we need to impose particular boundary conditions to the radial equation \eqref{final_radial} to simulate a scattering experiment. The boundary conditions for the angular equation \eqref{final_polar} should remain the same so that the scalar waves are regular at the polar boundaries of integration.

Assume an incident wave from the acceleration horizon, with amplitude coefficient $\mathcal{I}$, that scatters off the BH. In turn, the wave will be partially reflected back towards the acceleration horizon, with reflection coefficient $\mathcal{R}$, and partially transmitted through the potential barrier and into the event horizon, with transmission coefficient $\mathcal{T}$. The above translate to the radial boundary conditions \begin{equation}
	\label{scat}
	\psi \sim
	\left\{
	\begin{array}{lcl}
		\mathcal{T} e^{-i (\omega-\Phi(r_+))r_* },\,\,\,\,\,\quad\quad\quad\quad\quad\quad\,\, r \rightarrow r_+, \\
		&
		&
		\\
		\mathcal{I}e^{-i(\omega-\Phi(r_\alpha))r_*} + \mathcal{R} e^{i(\omega-\Phi(r_\alpha))r_*}, r \rightarrow r_\alpha.
	\end{array}
	\right.
\end{equation}
Due to the fact that the Wronskian of $\psi$ and its linearly-independent complex conjugate counterpart $\psi^\dagger$ does not depend on $r_*$, the Wronskians at both radial boundaries coincide, which subsequently leads to the relation
\begin{equation}\label{Wronskian}
	|\mathcal{R}|^2=|\mathcal{I}|^2-\frac{\omega-\Phi(r_+)}{\omega-\Phi(r_\alpha)}|\mathcal{T}|^2.
\end{equation}
From \eqref{Wronskian}, we observe that when the incident wave's frequency satisfies
\begin{equation}
	\label{suprad}
	\Phi(r_\alpha)<\omega<\Phi(r_+),
\end{equation}
then the amplitude of the reflected wave is larger than the amplitude of the incident one and the scalar wave is superradiantly amplified under the expense of the BH's electromagnetic energy. Notice that when $\alpha \rightarrow 0$, Eq. (\ref{suprad}) reduces to the respective superradiant condition for RN BHs \cite{Bekenstein:1973mi,Brito:2015oca}.

\begin{figure*}
	\includegraphics[scale=0.4]{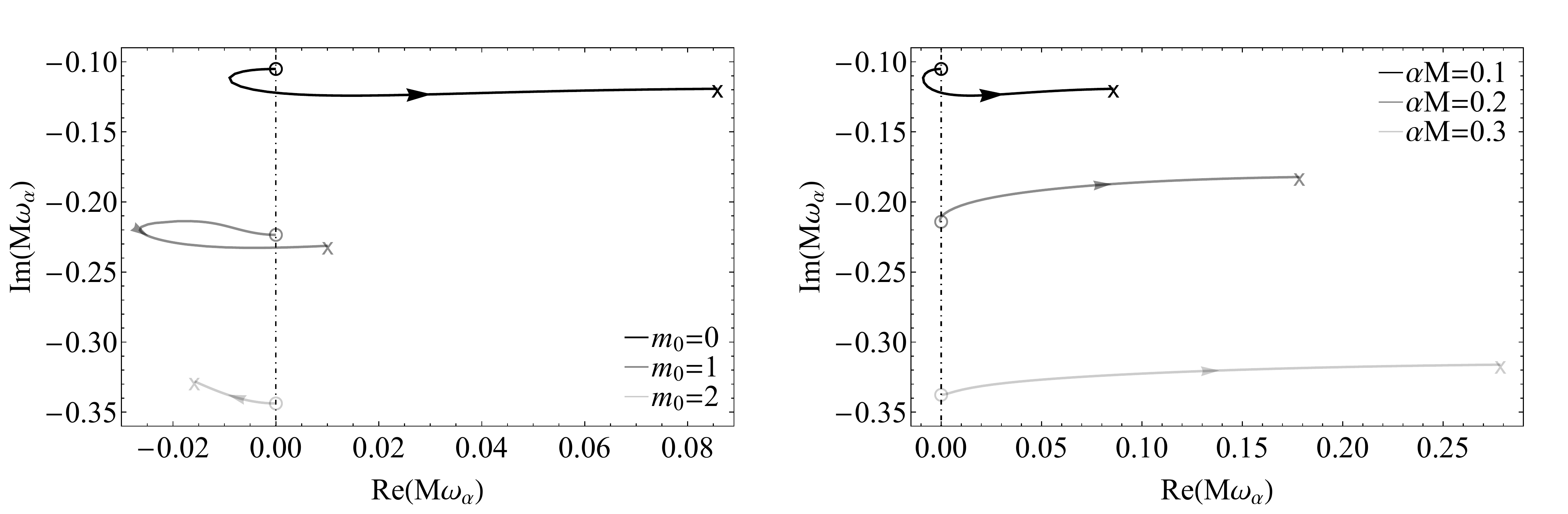}
	\caption{Left: Fundamental ($n=0$) charged scalar acceleration modes for an accelerating RN BH with $Q=0.5M$, $\alpha M=0.1$ and varying principal azimuthal number $m_0$. The charge coupling $qQ$ ranges from $0$ (circles) to $1$ (cross marks). Right: Fundamental ($n=0$) charged scalar acceleration modes with $m_0=0$ for an accelerating RN BH with $Q=0.5M$ and varying acceleration $\alpha M$. The charge coupling $qQ$ ranges from $0$ (circles) to $1$ (cross marks) following the direction of the arrows.}
	\label{accQNMs}
\end{figure*}

\begin{figure*}
	\includegraphics[scale=0.4]{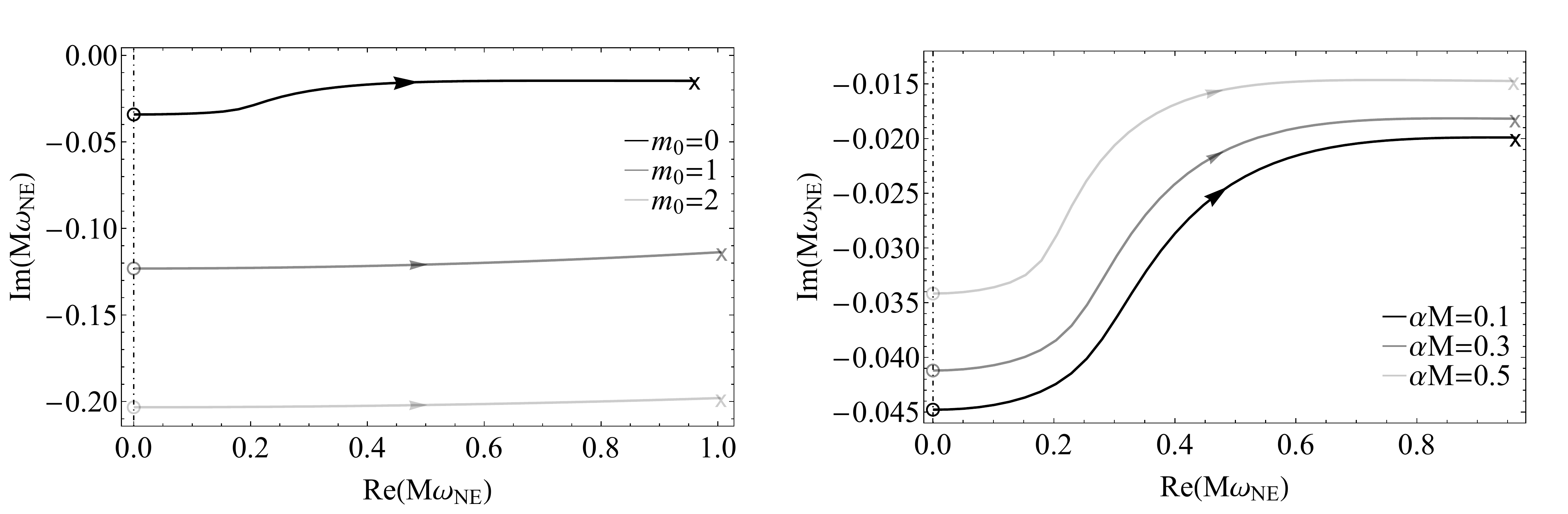}
	\caption{Left: Fundamental ($n=0$) charged scalar NE modes for an accelerating RN BH with $Q=0.999M$, $\alpha M=0.5$ and varying principal azimuthal number $m_0$. The charge coupling $qQ$ ranges from $0$ (circles) to $1$ (cross marks). Right: Fundamental ($n=0$) charged scalar NE modes with $m_0=0$ for an accelerating RN BH with $Q=0.999M$ and varying acceleration $\alpha M$. The charge coupling $qQ$ ranges from $0$ (circles) to $1$ (cross marks) following the direction of the arrows.}
	\label{NEQNMs}
\end{figure*}

To demonstrate our results, we define the amplification factor \cite{Brito:2015oca}
\begin{equation}
	Z_{m_0}=\frac{|\mathcal{R}|^2}{|\mathcal{I}|^2}-1,
\end{equation}
as a function of the incident wave's monochromatic frequency $\omega$ (not to be confused with the QNMs $\omega_n$). When $Z_{m_0}< 0$ superradiance does not occur while when $Z_{m_0}> 0$ the incident wave is superradiantly amplified with $Z_{m_0}=0$ defining the bounds of the superradiant relation \eqref{suprad}.

To numerically integrate Eq. \eqref{final_radial} with boundary conditions \eqref{scat}, we expand its solutions at the event and acceleration horizons to a desired order and match the two asymptotic solutions at an intermediate regime by imposing regularity of the solutions and their first derivatives. Through this numerical process we extract the amplification factor for numerous monochromatic incident waves with varying $\omega$. We have performed convergence tests by increasing the order of expansion of solutions at the boundaries of integration and observe that the amplification factors converge rapidly with the increment of the expansion order. In order to further validate our numerical results, we have compared the numerical onset and termination of superradiance by fitting our amplification factor curves to that designated by the exact relation \eqref{suprad} and have found excellent agreement.

\section{Results}

\subsection{Quasinormal modes of charged scalar fields}

The neutral scalar QNMs of accelerating Schwarzschild and RN BHs have been extensively analyzed in \cite{Destounis:2020pjk}. The dynamics of perturbations allows for three distinct families of solutions. The first family consists of the complex photon surface (PS) QNMs $\omega_\text{PS}$, which are intrinsically connected with the null geodesics' angular frequency and instability timescale at the equatorial plane of the photon surface and asymptote to the standard oscillatory QNMs of non-accelerating BHs when $\alpha\rightarrow0$ \cite{Cardoso:2008bp}. The acceleration modes $\omega_\alpha$, which are purely imaginary, linearly proportional to the acceleration parameter and vanish at $\alpha\rightarrow0$ correspond to another family completely independent of the PS QNMs. Finally, a last family, namely the near-extremal (NE) QNMs $\omega_\text{NE}$, consists of purely imaginary modes, depend on the surface gravity of the event (or Cauchy) horizon and only become relevant in the near-extremal limit $Q\rightarrow M$ of the charged $C$-metric. In this section, we extend the analysis of \cite{Destounis:2020pjk} to charged scalar QNMs in the available parameter space.

\begin{figure*}[]
	\includegraphics[scale=0.57]{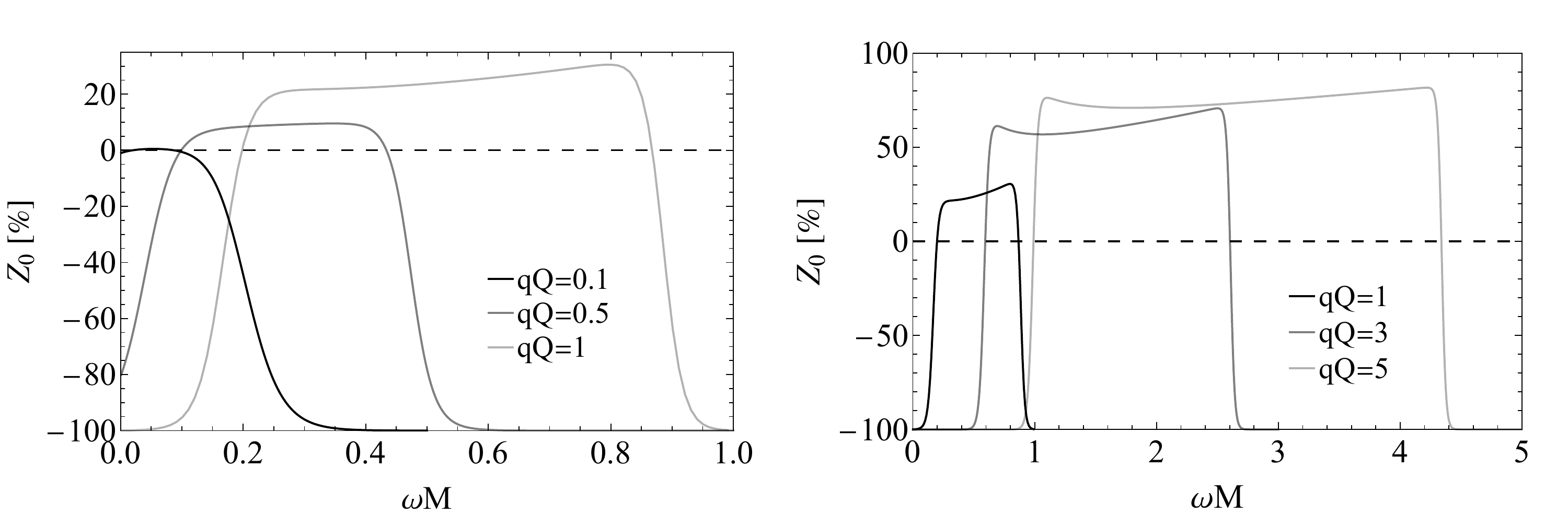}
	\caption{Amplification factors of massless charged $m_0=0$ monochromatic scalar waves for an accelerating RN BH with $\alpha M=0.2$, $Q=0.99M$ and varying charge coupling $qQ$. The horizontal black dashed line designates the onset of superradiant amplification.}
	\label{ampvsq}
\end{figure*}

\begin{figure}[t]
	\includegraphics[scale=0.57]{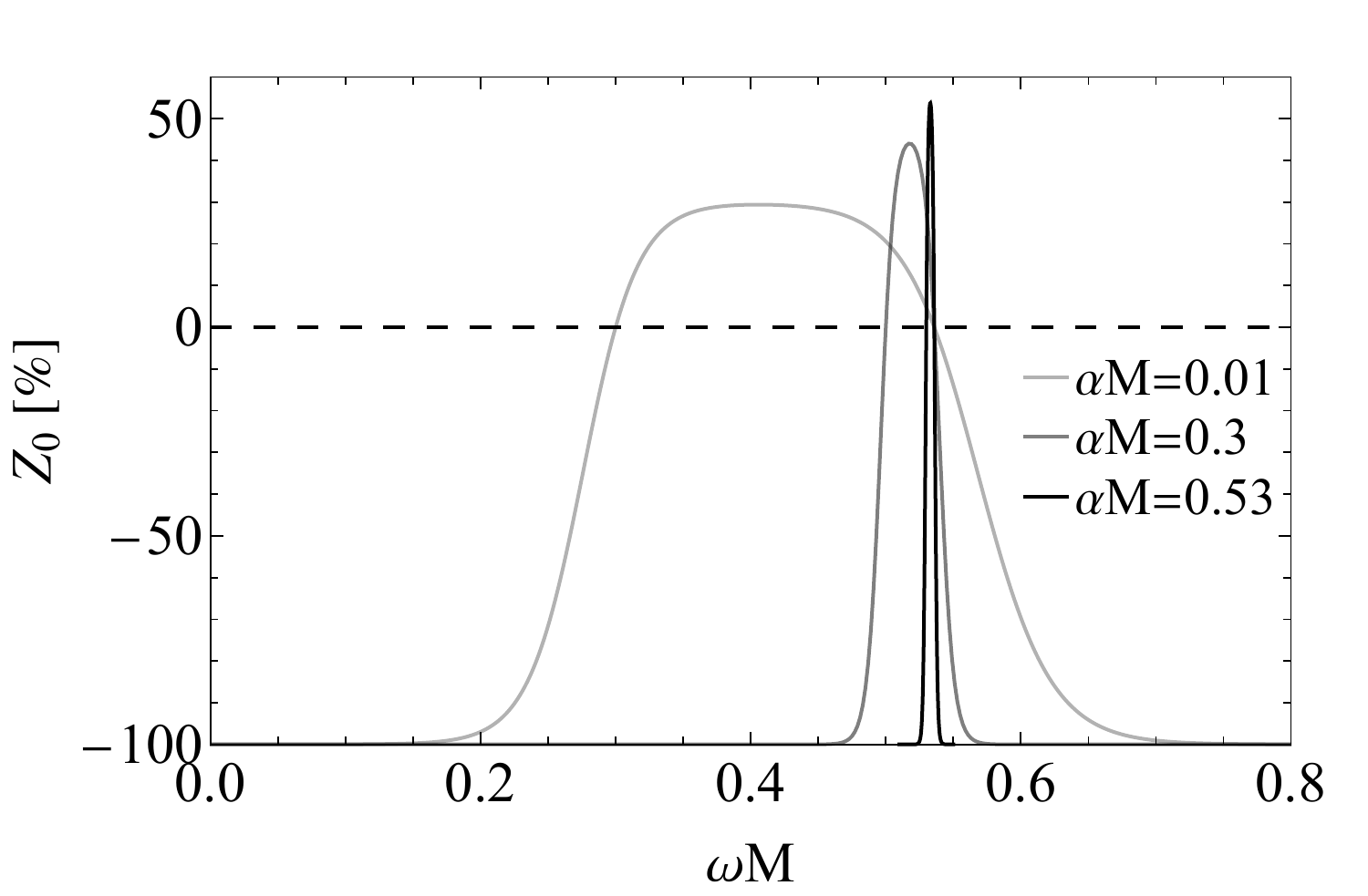}
	\caption{Amplification factors of massless charged $m_0=0$ monochromatic scalar waves with $qQ=1$ for an accelerating RN BH with $Q=0.5M$ and varying acceleration $\alpha M$. The horizontal black dashed line designates the onset of superradiant amplification.}
	\label{ampvsa05}
\end{figure}

\begin{figure*}
	\includegraphics[scale=0.57]{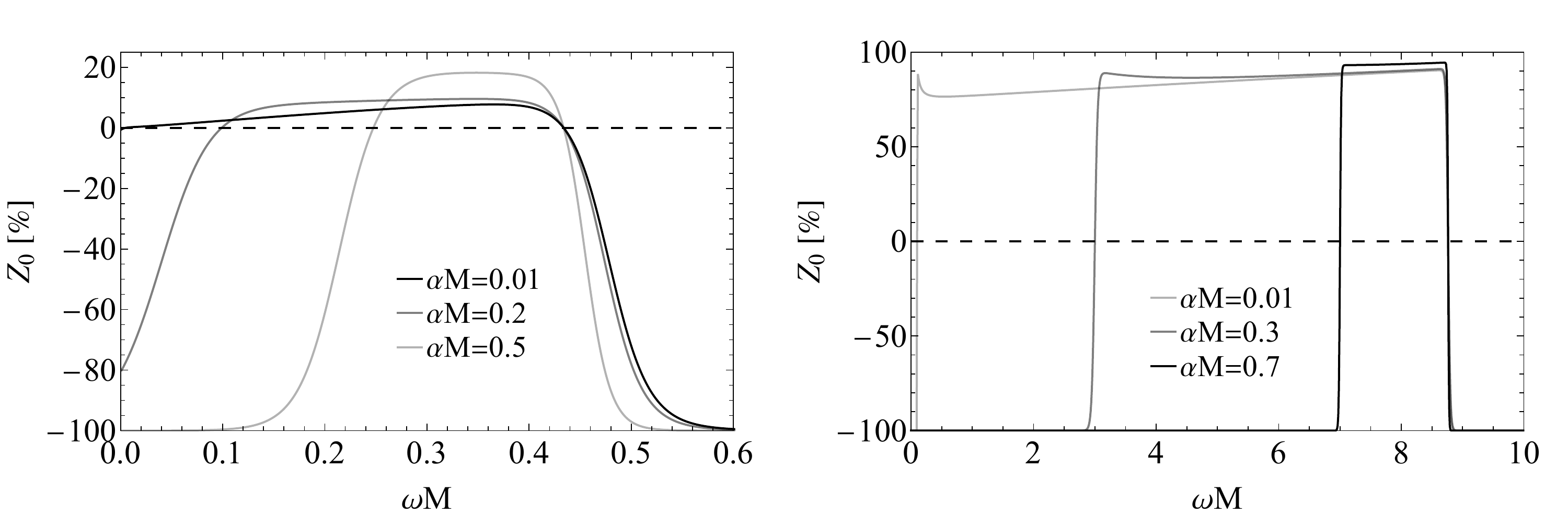}
	\caption{Left: Amplification factors of massless charged $m_0=0$ monochromatic scalar waves with $qQ=0.5$ for an accelerating RN BH with $Q=0.99M$ and varying acceleration parameter $\alpha M$. The horizontal black dashed line designates the onset of superradiant amplification. Right: Same as left with $qQ=10$.}
	\label{ampvsa}
\end{figure*}

Oscillatory QNMs, also referred as photon sphere modes \cite{Cardoso:2017soq,Destounis:2018qnb,Liu:2019lon,Destounis:2019omd}, share a symmetry with respect to reflections of the imaginary axis, that is the PS QNMs come in sets with common imaginary parts and opposite real parts. This degeneracy occurs due to the $\omega^2$ term in Eq. \eqref{final_radial}. The inclusion of charge to the scalar field breaks this symmetry, due to the introduction of a $\mathcal{O}(\omega)$ term in Eq. \eqref{final_radial} and a $qQ$-dependent term in the radial potential \eqref{pot_r}, to create two different branches of QNMs, which belong to the same family \cite{Cardoso:2018nvb}. A similar behavior is observed in accelerating RN BHs as depicted in Fig. \ref{PSQNMs}. The increment of $m_0$ leads to increased (in absolute value) oscillation frequencies and lifetime which is consistent with the behavior of Schwarzschild and RN QNMs when $\ell$ increases. The acceleration parameter has the same qualitative effect to QNMs as in the neutral case \cite{Destounis:2020pjk}, where its increment reduces the QNM's oscillation frequency and enlarges its lifetime. In contrast, the charge coupling $qQ$ interacts with the effects of $m_0$ and $\alpha M$ by increasing the frequency and decay rate of the positive branch (solid curves in Fig. \ref{PSQNMs}) while increasing the frequency and decreasing the decay rate of the negative branch (dashed curves in Fig. \ref{PSQNMs}). Similar branching behavior has been found in Kerr and Kerr-Newman BHs \cite{Onozawa:1996ux,Berti:2004md,Berti:2005eb,Vieira:2021ozg}.

The purely imaginary acceleration modes become complex when the charge coupling is introduced, in similarity to the de Sitter QNMs of RNdS BHs \cite{Destounis:2019hca,Destounis:2018qnb}. Figure \ref{accQNMs} depicts the particulars of the acceleration QNMs' migration in the complex plane when $qQ>0$. Initially, as $qQ$ increases, the family obtains a negative real part which reaches a turning point beyond which larger $qQ$ values increase the real part and eventually turn it positive. Such behavior occurs due to the competition between the $\mathcal{O}(\omega)$ term and the radial potential in Eq. \eqref{final_radial}. The turning point depends on $m_0$, with larger $m_0$ having deeper turning points in the complex plane region with $\text{Re}(M\omega_\alpha)<0$, and $\alpha M$, with its increment saturating the turning point. The decay timescales have the general tendency to remain around the vicinity of the purely imaginary acceleration modes, for which $qQ=0$, for small $m_0$ and sufficiently large $\alpha M$. 

Analogously, the purely imaginary NE modes obtain an oscillation frequency when the charge coupling is positive, in a similar fashion to the NE RNdS QNMs \cite{Cardoso:2018nvb,Destounis:2018qnb}. In Fig. \ref{NEQNMs} we demonstrate their behavior with respect to $m_0$ and $\alpha M$ as $qQ$ increases. The modes branch from the imaginary axis with oscillation frequencies loosely proportional to $qQ$. The effect of $m_0$ and $\alpha M$ is identical to that observed in \cite{Destounis:2020pjk}, where the increment of $m_0$ ($\alpha M$)  decreases (increases) their lifetime. Nevertheless, as the charge coupling becomes stronger, the lifetime of NE QNMs is increased till it reaches a plateau, in full agreement with the NE modes of RNdS geometry \cite{Cardoso:2018nvb,Destounis:2018qnb}.

Throughout the overall QNM analysis, we have not distinguished any unstable QNMs, therefore we conjecture that accelerating RN BHs are modally stable against linear charged scalar fluctuations. Furthermore, we did not recover QNMs at any regime of the available parameter space that satisfy the supperadiant relation \eqref{suprad}, hence we should not expect resonant behavior in the amplification factors of scattered charged scalar waves on accelerating BHs. This does not guarantee that the phenomenon of superradiance does not occur in such spacetimes (we tackle this investigation on the following subsection). Nonetheless, these results contrast those found in RNdS BHs, where charged scalar perturbations are superradiantly unstable \cite{Zhu:2014sya,Konoplya:2014lha,Destounis:2019hca} and lead to resonant hyperradiation with the amplification factors reaching very high levels close to superradiant QNMs \cite{Mascher:2022pku}.

\subsection{Superradiant amplification of charged scalar waves}

\begin{figure}[]
	\includegraphics[scale=0.57]{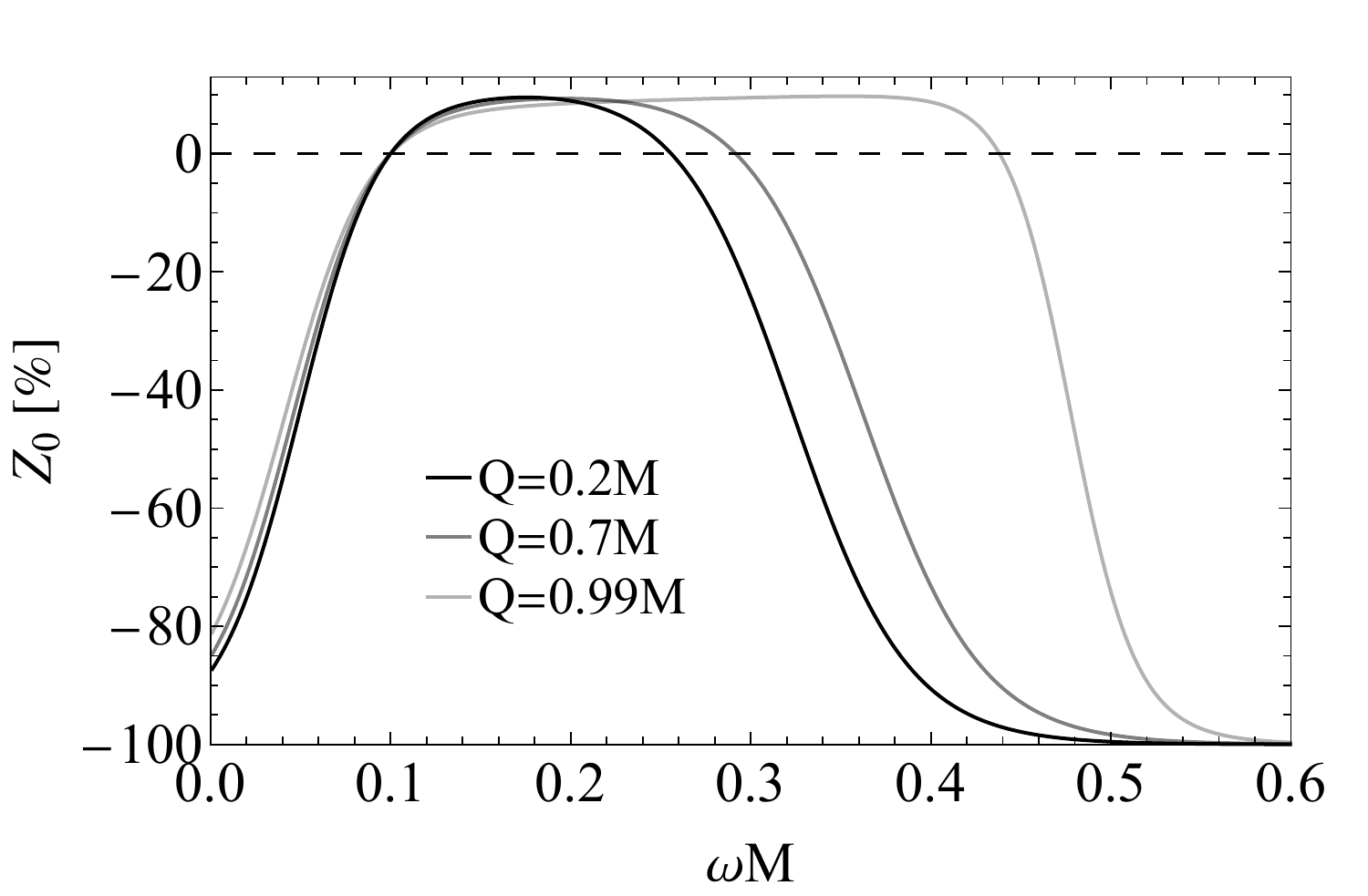}
	\caption{Amplification factors of massless charged $m_0=0$ monochromatic scalar waves with $qQ=0.5$ for an accelerating RN BH with $\alpha M=0.2$ and varying electric charge $Q/M$. The horizontal black dashed line designates the onset of superradiant amplification.}
	\label{ampvsQ}
\end{figure} 

\begin{figure}[]
	\includegraphics[scale=0.57]{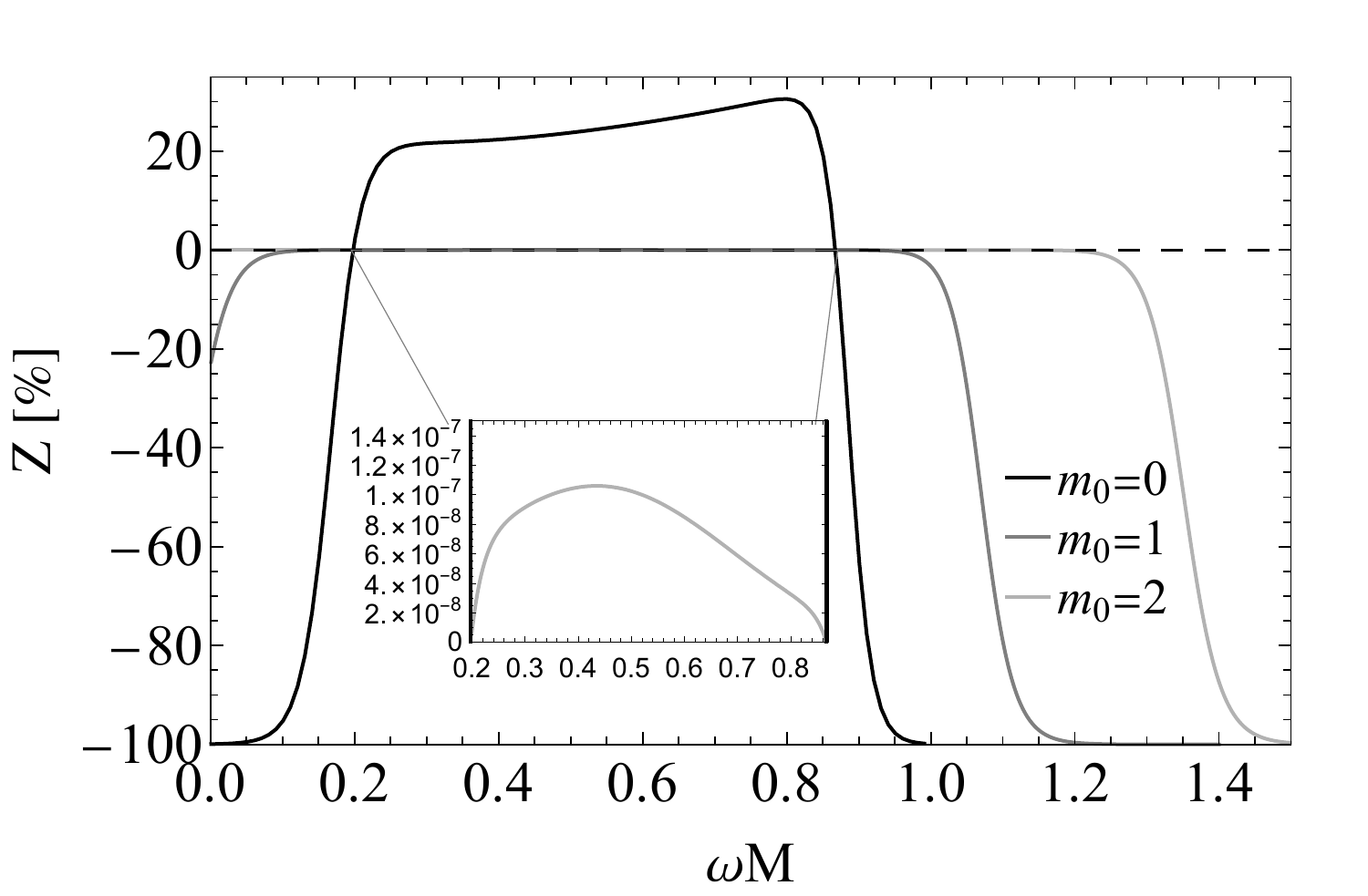}
	\caption{Amplification factors of massless charged monochromatic scalar waves with $qQ=1$ and varying $m_0$ for an accelerating RN BH with $\alpha M=0.2$ and $Q=0.99M$. The horizontal black dashed line designates the onset of superradiant amplification.}
	\label{ampvsm}
\end{figure}

RN BHs are known to favor the amplification of charged incident monochromatic scalar waves under the expense of their electromagnetic energy \cite{Bekenstein:1973mi,Brito:2015oca}. In this section, we extend such investigation in the charged $C$-metric in order to interpret the effects of acceleration in the phenomenon of supperadiant scattering. 

Figure \ref{ampvsq} demonstrates that an accelerating charged BH amplifies incident monochromatic charged scalar waves. At the large scalar charge limit $qQ>>1$ the amplification tends to $100\%$, in similarity with RN BHs. The inclusion of acceleration induces a lower bound for the onset of superradiance, in contrast to RN BHs which can amplify incident waves with arbitrarily small frequencies \cite{Brito:2015oca}. Most importantly, increasing $\alpha M$ leads to further amplification of incident waves, with respect to RN, even though the superradiant frequency range is significantly shortened close to extremal accelerations, i.e. $\alpha \rightarrow1/r_+$ (see Fig. \ref{ampvsa05}). In Fig. \ref{ampvsa} we confirm that a favorable combination of acceleration and charge coupling maximizes the amplification such that $Z_0\rightarrow 100\%$, though our numerical investigation entails that the charge coupling is more efficient in the maximization of superradiance than the acceleration. Nevertheless, the increment of $\alpha M$ diminishes the frequency range for which accelerating BHs superradiate, in similarity with Kerr-dS \cite{Tachizawa:1992ue} and RNdS BHs \cite{Mascher:2022pku}.

Increasing the BH charge does not seem to play an important role on the amplification, though it enlarges the frequency domain where superradiance occurs (see Fig. \ref{ampvsQ}). In finality, Fig. \ref{ampvsm} demonstrates the expected effect of $m_0$ which rapidly diminishes superradiance when increased and is the fundamental reason why we mainly focused our analysis on $m_0=0$ charged scalar waves. Since we have not found any superradiant QNMs in the previous subsection, we did not expect to spot resonant peaks in the amplification factors and all figures conform with this result.

\section{Conclusions}

The charged $C$-metric is a boost-symmetric spacetime which describes charged BHs that are uniformly accelerated under the tension of a cosmic string. The acceleration horizon that is generally present in such spacetimes shares many similarities with the cosmological horizon of de Sitter BHs and the overall causal structure of the charged $C$-metric resembles that of RNdS geometries. Due to the existence of QNM instabilities \cite{Zhu:2014sya,Konoplya:2014lha,Destounis:2019hca}, as well as the presence of superradiant hyperradiation in RNdS BHs against charged scalar fields \cite{Mascher:2022pku}, an analogous investigation of these phenomena in charged accelerating BHs is of interest. 

In this work, we have extended the neutral scalar QNM analysis of \cite{Destounis:2020pjk} to charged scalar perturbations and initiated the study of superradiance in charged accelerating BHs. In spite of their resemblance with RNdS spacetime, we did not obtain any superradiantly unstable charged scalar QNMs, thus we conjecture that the charged $C$-metric is modally stable against charged scalar fluctuations. The introduction of scalar charge breaks the reflection symmetry of PS modes with respect to the imaginary axis and leads to a non-trivial migration of both QNM branches in the complex plane. In turn, the acceleration and NE families of purely imaginary modes acquire an oscillation frequency when a scalar charge is introduced and never asymptote to each other at any limit. 

From the behavior of the NE modes and the absence of superradiantly unstable modes we can conjecture that the violation of strong cosmic censorship is imminent in the charged $C$-metric when charged scalar fields are considered, in analogy with \cite{Cardoso:2018nvb}, though the increment of the scalar charge should alleviate these violations if `wiggles' of the NE family's imaginary part, such as those found in \cite{Destounis:2018qnb}, are not present in this spacetime. A more precise analysis would shed more light into the deterministic nature of GR regarding charged $C$-metric geometries.

The absence of QNMs which satisfy the superradiant condition certainly conveys that the amplification of incident monochromatic charged scalar waves should resemble that occurring in RN BHs. More specifically, such superradiant amplification from accelerating RN BHs should be similar to that of Kerr-de Sitter BHs \cite{Tachizawa:1992ue} due to the existence of the acceleration horizon which imposes a lower bound for the onset of superradiance. Indeed, our investigation agrees qualitatively with the superradiant amplification of massless scalar waves in Kerr-de Sitter spacetime with a twist; the increment of the charge coupling $qQ$ can push the amplification factors arbitrarily close to $100\%$. Furthermore, the increment of the acceleration parameter shortens the superradiant frequency range though leads to the elevation of amplification and can also reach to $100\%$ near the extremal acceleration. Therefore, charged accelerating BHs are more efficient deposits of energy than RN and Kerr-de Sitter BHs, though one has to hypothesize the existence of cosmic strings in order to provide the perpetual energy needed for BHs to accelerate.

A promising direction to further elucidate these compact objects would be to extend the current analysis in order to include a cosmological constant to the spacetime and/or a mass to the charged scalar field. Even so, such investigation requires an immense amount of work due to the fact that the Klein-Gordon equation for massive scalar fields is not conformally-invariant anymore and even for massless fields, the inclusion of a cosmological constant requires a much more intricate conformal transformation for the Klein-Gordon equation to be conformally-invariant (see \cite{Charmousis:2009cm,Nozawa:2008wf,Fontana:2022whx}). 

\begin{acknowledgments}
	The authors would like to warmly thank Rodrigo Fontana and Filipe Mena for helpful discussions.
\end{acknowledgments}

\appendix

\section{Conformal transformations}\label{appA}

One of the most challenging aspects of the charged $C$-metric is the presence of the scale factor $\Omega^{-2}$ which multiplies all non-vanishing terms of the metric tensor in Eq. \eqref{Cmetric}. Even though this factor introduces a coupling between $r$ and $\theta$, which eventually ruins a potential separation of variables, there is a particular way of rescaling a spacetime through conformal transformations in order to achieve separability of perturbation equations. Conformal transformations occur in many contexts in gravity, e.g. to define the asymptotic structure of spacetime \cite{Penrose:1962ij,Penrose:1965am,Penrose:1968ar} and to relate different representations of modified theories of gravity \cite{Dicke:1961gz,Harrison:1973kb}. In what follows, we collect the appropriate conformal transformation relations regarding scalars, vectors and tensors that appear in the charged Klein-Gordon equation \eqref{conformal_tilde_wave_equation_4d}.

A conformal transformation is a position-dependent rescaling of a spacetime metric $g_{\mu\nu}$, so that $\tilde{g}_{\mu\nu}=\Omega^2 g_{\mu\nu}$, where the conformal factor $\Omega(x^\mu)$ is a dimensionless positive smooth function of the spacetime position four-vector $x^\mu$. Conformal transformations do not change the metric signature, the sign of the magnitude of the four-vectors, the angles between them and, most importantly, they leave the light cones and the causal structure of spacetime invariant \cite{Wald-book}. By denoting the inverse of $g_{\mu\nu}$ as $g^{\mu\nu}$ and the inverse of the conformally-rescaled metric $\tilde{g}_{\mu\nu}$ as $\tilde{g}^{\mu\nu}$ we can derive the following:
\begin{equation}
	\tilde{g}^{\mu\nu}\tilde{g}_{\nu\rho}=\Omega^s g^{\mu\nu}\Omega^2 g_{\nu\rho}=\Omega^{s+2}\delta^\mu_\rho,
\end{equation}
where $s\in \mathbb{R}$ is the conformal weight. For $\tilde{g}^{\mu\nu}\tilde{g}_{\nu\rho}=\delta^\mu_\rho$ to hold then, $s=-2$, thus $\tilde{g}^{\mu\nu}=\Omega^{-2}g^{\mu\nu}$. If we denote $\nabla_\mu$ and $\tilde{\nabla}_\mu$ the covariant derivative operators associated with $g_{\mu\nu}$ and $\tilde{g}_{\mu\nu}$, respectively, then the relation between the two operators is $\tilde{\nabla}_\mu V_\nu=\nabla_\mu V_\nu-C^\rho_{\mu\nu}V_\rho$, where \cite{Wald-book,Carroll:2004st}
\begin{align}
	\tilde{\Gamma}^\rho_{\mu\nu}&={\Gamma}^\rho_{\mu\nu}+C^\rho_{\mu\nu},\\
	C^\rho_{\mu\nu}&=\frac{1}{2}\tilde{g}^{\rho\kappa}\left(\nabla_\mu\tilde{g}_{\nu\kappa}+\nabla_\nu\tilde{g}_{\mu\kappa}-\nabla_\kappa\tilde{g}_{\mu\nu}\right).
\end{align}
Notice that when $\tilde{\nabla}_\mu$ is the standard covariant derivative, then ${\nabla}_\mu=\partial_\mu$ and $C^\rho_{\mu\nu}$ are the well-known Christoffel symbols $\Gamma^\rho_{\mu\nu}$. However, since $\nabla_\mu g_{\nu\rho}=0$, we have
\begin{equation}\label{nabla}
	\nabla_{\mu}\tilde{g}_{\nu\rho}=\nabla_\mu\left(\Omega^2 g_{\nu\rho}\right)=2\Omega g_{\nu\rho} \nabla_\mu \Omega.
\end{equation}
Using Eq. \eqref{nabla}, we can express $C^\rho_{\mu\nu}$ in terms of $g_{\mu\nu}$ and $\Omega$ as
\begin{align}
	C^\rho_{\mu\nu}&=\Omega^{-1}g^{\rho\kappa}\left(g_{\nu\kappa}\nabla_\mu\Omega+g_{\mu\kappa}\nabla_\nu\Omega-g_{\mu\nu}\nabla_\kappa\Omega\right),\nonumber\\\label{C}
	&=\delta^\rho_\nu\nabla_\mu\ln\Omega+\delta^\rho_\mu\nabla_\nu\ln\Omega-g^{\rho\kappa}g_{\mu\nu}\nabla_\rho\ln\Omega.
\end{align}

In what follows we will need the covariant derivative of a scalar field $\phi$ and the Ricci curvature $R\equiv g^{\mu\nu}R_{\mu\nu}$, where $R_{\mu\nu}$ is the Ricci tensor. The covariant derivatives of scalar fields are equal in both the original and conformal frames, since they are both equal to the partial derivative $\tilde{\nabla}_\mu\phi=\nabla_\mu\phi=\partial_\mu\phi$. The Ricci tensor in the conformal frame reads \cite{Wald-book,Carroll:2004st}
\begin{align}\label{ricci}
	\tilde{R}&=\Omega^{-2}\left(R-2(N-1)g^{\mu\rho}\nabla_\mu\nabla_\rho\ln\Omega\right.\nonumber\\
	&\left.-(N-2)(N-1)g^{\mu\rho}\left[\nabla_\mu\ln\Omega\right]\nabla_\rho\ln\Omega\right),
\end{align}
where $N$ is the spacetime dimension.

Finally, we require knowledge of conformal transformations of the electromagnetic field equations. Maxwell's equations are invariant under conformal transformation in four dimensions \cite{Wald-book,Cote:2019kbg}. Physically, this is true due to the fact that the photon is massless and therefore no length or mass scale is associated with the electromagnetic field. Assuming that the conformal electromagnetic tensor, associated with the metric $\tilde{g}_{\mu\nu}$, is $\tilde{F}_{\mu\nu}=\Omega^s F_{\mu\nu}$, where $s$ is, again, a conformal weight, then for the non-homogeneous Maxwell's equations $\nabla^\mu F_{\mu\nu}=-4\pi J_\nu$, $\nabla_{[\mu}F_{\nu\rho]}=0$, with $J_\mu$ a four-current, to be conformally invariant in four dimensions, the conformal weight should be $s=0$ \cite{Wald-book,Cote:2019kbg}. Thus, 
\begin{equation}\label{Maxwell}
	\tilde{F}_{\mu\nu}=F_{\mu\nu}, \quad \tilde{F}^{\mu\nu}=\Omega^{-4}F^{\mu\nu}, \quad \tilde{J}_\mu=\Omega^{-2}J_\mu. 
\end{equation}
The validity of Maxwell's equations in the conformally-rescaled spacetime guarantees that 
\begin{equation}
	\tilde{F}_{\mu\nu}=\tilde{\nabla}_\mu \tilde{A}_\nu-\tilde{\nabla}_\nu\tilde{A}_\mu=\partial_\mu\tilde{A}_\nu-\partial_\nu\tilde{A}_\mu,
\end{equation}
where $A_\mu$ is the electromagnetic four-potential. If we assume $\tilde{A}_\mu=\Omega^s A_\mu$, then
\begin{equation}
	\tilde{F}_{\mu\nu}=s \Omega^{s-1}\left(\partial_\mu\Omega A_\nu-\partial_\nu\Omega A_\mu\right)+\Omega^s F_{\mu\nu},
\end{equation}
and since $\tilde{F}_{\mu\nu}=F_{\mu\nu}$ then
\begin{equation}
	\left(\Omega^s-1\right) F_{\mu\nu}+s \Omega^{s-1}\left(\partial_\mu\Omega A_\nu-\partial_\nu\Omega A_\mu\right)=0,
\end{equation}
which is satisfied if $s=0$. Therefore,
\begin{equation}\label{covpot}
	\tilde{A}_\mu=A_\mu,\quad \tilde{A}^\mu=\Omega^{-2}A^\mu.
\end{equation}

\section{Conformal invariance of the charged scalar wave equation}\label{appB}

The Klein-Gordon equation for a scalar field $\phi$ is said to be conformally-invariant if there exists a conformal weight $s$ such that $\phi$ is a solution with metric $g_{\mu\nu}$ if and only if $\tilde{\phi}=\Omega^s\phi$ is a solution with metric $\tilde{g}_{\mu\nu}=\Omega^2 g_{\mu\nu}$.
Here, we show that the Klein-Gordon equation for charged scalar fields in curved spacetime 
\begin{equation}\label{wave equation}
	g^{\mu\nu}D_\mu D_\nu\phi=0,
\end{equation}
is not conformally invariant if the spacetime dimension satisfies $N\neq 2$. By using the conformal transformations of Appendix \ref{appA}, we have
\begin{align}
	&\tilde{g}^{\mu\nu}\tilde{D}_\mu\tilde{D}_\nu\tilde{\phi}=\Omega^{-2}g^{\mu\nu}\left(\tilde{\nabla}_\mu-iq\tilde{A}_\mu\right)\left(\tilde{\nabla}_\nu-iq\tilde{A}_\nu\right)\left(\Omega^s\phi\right)\nonumber\\
	&=\Omega^{-2}g^{\mu\nu}\tilde{\nabla}_\mu\tilde{\nabla}_\nu\left(\Omega^s\phi\right)-iq\left(2s+N-2\right)\phi\Omega^{s-3}g^{\mu\nu}A_\mu\nabla_\nu\Omega\nonumber\\\label{cov}
	&-2iq\Omega^{s-2}g^{\mu\nu}A_\mu\nabla_\nu\phi-q^2\Omega^{s-2}\phi g^{\mu\nu}A_\mu A_\nu
\end{align}
where the first term of Eq. \eqref{cov} is \cite{Wald-book}
\begin{align}
	&\Omega^{-2}g^{\mu\nu}\tilde{\nabla}_\mu\tilde{\nabla}_\nu\left(\Omega^s\phi\right)=\Omega^{s-2}g^{\mu\nu}\nabla_\mu\nabla_\nu\phi\nonumber\\
	&+(2s+N-2)\Omega^{s-3}g^{\mu\nu}\nabla_\mu\Omega\nabla_\nu\phi+s \Omega^{s-3}\phi g^{\mu\nu}\nabla_\mu\nabla_\nu\Omega\nonumber\\\label{neutral}
	&+s(s+N-3)\Omega^{s-4}\phi g^{\mu\nu}\nabla_\mu\Omega\nabla_\nu\Omega,
\end{align}
and Eqs. \eqref{C} and \eqref{covpot} were used. If $N=2$, we may choose $s=0$ so that Eq. \eqref{cov} becomes
\begin{align}
	\tilde{g}^{\mu\nu}\tilde{D}_\mu\tilde{D}_\nu\tilde{\phi}&=\Omega^{-2}\left(g^{\mu\nu}\nabla_\mu \nabla_\nu\phi-2iqg^{\mu\nu}A_\mu\nabla_\nu\phi\right.\nonumber\\&\left.-q^2\phi g^{\mu\nu}A_\mu A_\nu\right)\nonumber\\
	&=\Omega^{-2}g^{\mu\nu}D_\mu D_\nu\phi=0,
\end{align}
where the Lorenz gauge $\nabla_\mu A^\mu=0$ was used. Therefore,
$\tilde{g}^{\mu\nu}\tilde{D}_\mu\tilde{D}_\nu\tilde{\phi}=0$ holds if and only if $g^{\mu\nu}D_\mu D_\nu\phi=0$ in two-dimensional spacetimes. However, if $N\neq 2$, there is no choice for $s$ which makes the Klein-Gordon equation conformally invariant.

Even so, for $N>1$ it is possible to modify Eq. \eqref{cov} so that it becomes conformally invariant. If we choose $s=1-N/2$, then the $\nabla_{\mu}\Omega\nabla_{\nu}\phi$ terms in \eqref{neutral} and the $A_\mu\nabla_{\nu}\Omega$ term in \eqref{cov} are eliminated since the prefactor $(2s+n-2)=0$. By fixing $s=1-N/2$, we can further modify the Klein-Gordon equation \eqref{cov} with an additional term of the form $\beta \tilde{R} \phi$, where $\beta=-(N-2)/(4(N-1))$. By doing so, the additional terms of the Ricci scalar in the conformal frame \eqref{ricci} will cancel out the $\phi g^{\mu\nu}\nabla_{\mu}\nabla_{\nu}\Omega$ and $\phi g^{\mu\nu}\nabla_{\mu}\Omega\nabla_{\nu}\Omega$ terms in \eqref{neutral}. Therefore, the modified Klein-Gordon equation for charged scalar fields (also known as the covariant wave equation)
\begin{equation}\label{conformal_wave_equation}
	g^{\mu\nu}D_\mu D_\nu\phi-\frac{N-2}{4(N-1)}R\phi=0,
\end{equation}
is conformally invariant, when $s=1-N/2$ since
\begin{align}
	&\left(\tilde{g}^{\mu\nu}\tilde{D}_\mu \tilde{D}_\nu-\frac{N-2}{4(N-1)}\tilde{R}\right)\left[\Omega^{1-N/2}\phi\right]=\nonumber\\
	&\Omega^{-1-N/2}\left(	g^{\mu\nu}D_\mu D_\nu-\frac{N-2}{4(N-1)}R\right)\phi.
\end{align}
Thus, Eq. \eqref{conformal_wave_equation} provides a conformally-invariant generalization of the Klein-Gordon equation for charged scalar fields to curved geometries.

\bibliography{C_metric}

\end{document}